# Computation of K-Core Decomposition on Giraph

Fangming Liu (jasmine1@uvic.ca)
Supervisor: Prof. Alex Thomo
University of Victoria
Spring 2017

## Abstract

Graphs are an essential data structure that can represent the structure of social networks. Many online companies, in order to provide intelligent and personalized services for their users, aim to comprehensively analyze a significant amount of graph data with different features. One example is k-core decomposition which captures the degree of connectedness in social graphs. The main purpose of this report is to explore a distributed algorithm for k-core decomposition on Apache Giraph. Namely, we would like to determine whether a cluster-based, Giraph implementation of k-core decomposition that we provide is more efficient than a single-machine, disk-based implementation on GraphChi for large networks. In this report, we describe (a) the programming model of Giraph and GraphChi, (b) the specific implementation of k-core decomposition with Giraph, and (c) the result comparison between Giraph and GraphChi. By analyzing the results, we conclude that Giraph is faster than GraphChi when dealing with large data. However, since worker nodes need time to communicate with each other, Giraph is not very efficient for small data.

## Report Specification

## Figures

## Tables

## 1. Introduction

Graphs are popularly used in representing connections between people or entities. For example, graphs play a significant role in modeling social networks. Each vertex in the graph can be treated as an individual, and each edge between vertices shows the relationships between individuals. By analyzing the data in the graph, a social networks company could know the specific social circle for each user, so that it can recommend some new friends to the user based on user's current friends. Similarly, graphs are also widely used in shopping websites. Clients can receive their expected ads according to the websites that they often visit. To consider these problems further, what if a user unfollowed another user, or a client disliked some products he previously liked? After these interactions, the user's friends may also unfollow the same user, and the client may also dislike the similar products that he wanted before. In this case, the sparse distribution of the current graph might be dramatically changed. This is related to a graph concept, k-core, which measures how sparse (or how well connected) a graph is. The k-core of a graph represents the maximal subgraph in which every vertex is connected to at least k vertices in the subgraph. k-core is used for community detection, protein function prediction, visualization, and solving NP-hard problems on real networks. Therefore, it is an important concept in graph data analytics.

In this work, we use Apache Giraph to implement a distributed vertex-centric algorithm by Montresor et. al. [22]. The largest dataset we used in this report has around 4.8 million nodes and 69 million edges. Our goal is not only to examine our implementation on Giraph, but also compare it to an implementation provided by Khaouid et. al. [17] on GraphChi, a disk-based vertex-centric system [19].

## 2. Vertex-Centric Computation

### 2.1 Giraph Programming Model

Giraph provides a programming model and API for users, which makes users just need to focus on implementing user-defined functions (UDF) for each vertex. Figure 2.1 shows this conceptual organization of an application in Giraph. Users do not need to be concerned about graph storage, algorithms execution, or computation distribution of the slaves. Additionally, they do not even need to worry about how to iterate their UDF on each vertex, since Giraph will automatically go through every active vertex in each superstep.

The key point of a Giraph implementation is to write the proper UDF to manage the behaviour of vertices in the graph. During the computation, each vertex can be either active or inactive. As Figure 2.2 shows below, all of the vertices are in active state at the beginning of the computation. A vertex will vote to halt when its work is done, and its status will be switched from active to inactive. If a vertex receives a message in some subsequent superstep, it would become active again. The whole computation process finishes when all of the vertices are in the inactive state, and there is no vertex that needs to send any message.

The computation process proceeds as follows. In the beginning of the computation, the master machine assigns a portion of the graph to each slave machine. Slaves synchronously compute their loaded vertices, and the result is aggregated from each slave. Slaves continue to compute the active vertices and the results are aggregated again.

### 2.2 GraphChi Programming Model

Similar to the Giraph model, GraphChi also offers a programming model and API for users, so that users just need to implement the update function for the vertices in the graph. Compared to the Giraph model, the main difference is that GraphChi is a disk-based model. Instead of using multiple slaves, it only needs a single machine to implement the whole calculation. Additionally, GraphChi can customize the scheduling vertices. Users can selectively schedule the vertices that need to be updated to save the running time of the computation.

## 3. Giraph Implementation

To explore the relationship between the number of the slaves and the running time, we respectively use two, five, ten, fifteen, and twenty slaves to handle different datasets. The datasets we used in this experiment are chosen from Stanford Large Network Dataset Collection. They are Astro Physics (ca-AstroPh), Gnutella P2P network (p2p-Gnutella31), Amazon product co-purchasing network (amazon0601), California road network (roadNet-CA), and Liver-Journal social network (soc-LiveJournal1). The detailed information of these datasets is described in Table 3.1. From the table, we can observe that the first two datasets are small; they only have few thousands of nodes and a hundred thousands of edges. The medium sized datasets are amazon0601 and roadNet-CA with around three million edges. The largest dataset is soc-LiveJournal1, which has 4.8 million nodes and approximately 69 million edges.

| Dataset Name | Numbers of Nodes | Numbers of Edges |
|---|---|---|
| ca-AstroPh | 18,772 | 198,110 |
| p2p-Gnutella31 | 62,586 | 147,892 |
| amazon0601 | 403,394 | 3,387,388 |
| roadNet-CA | 1,965,206 | 2,766,607 |
| soc-LiveJournal1 | 4,847,571 | 68,993,773 |

Table 3.1 Datasets Information

## 4. Setting up Apache Giraph on Amazon Web Services (AWS) EC2

The experiment is conducted on Amazon Web Services (AWS) using Amazon Elastic Compute Cloud (EC2) platform. We configured twenty-one virtual machines, with one master machine and twenty slaves. All of the virtual machines have two cores, Intel Xeon Family, 2.4 GHz CPU with 8GB RAM running Ubuntu Linux System. The specific steps of instance configuration and Giraph implementation on EC2 platform are described below.

### 4.1.1 Launch Instances

1) Go to the Amazon EC2 Website: *https://aws.amazon.com/ec2/*, and click on "Create an AWS Account" in the upper-right corner.

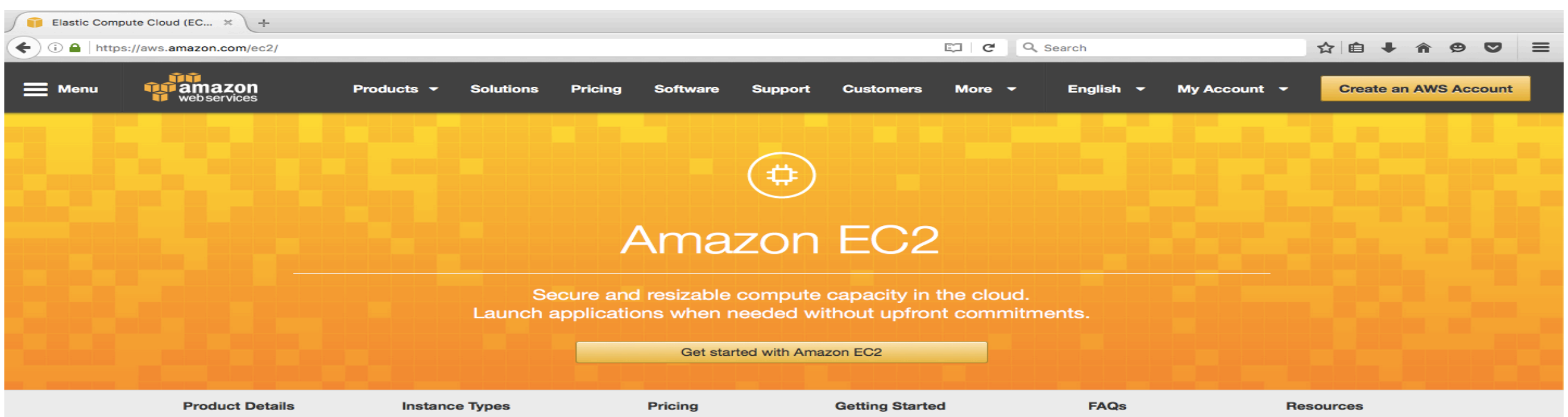

Figure 3.1 Launch Instances (1)

2) After the account is created successfully, we will navigate to the following page. We need EC2 services to conduct our experiment, so click on "EC2" in the first column.

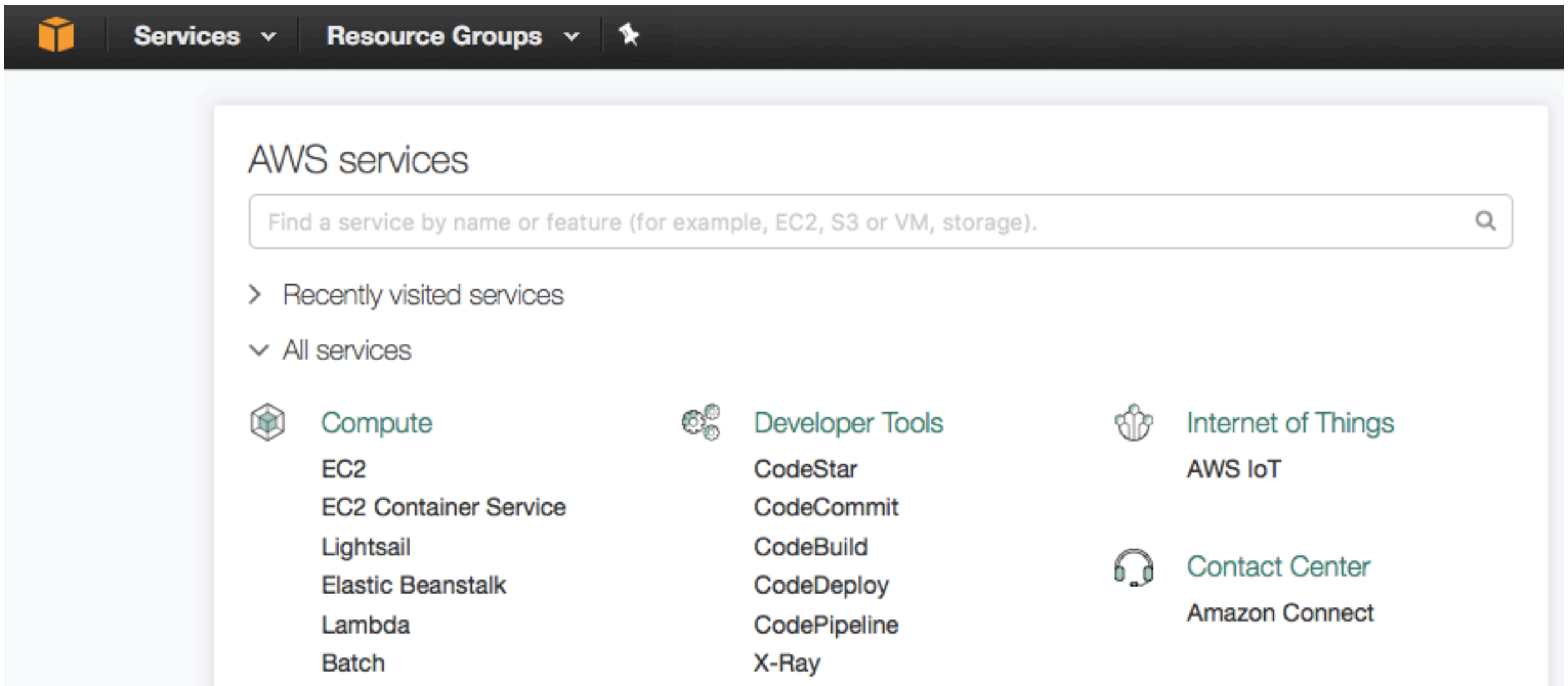

Figure 3.2 Launch Instances (2)

3) Click on "Launch Instance" to configure the instances.

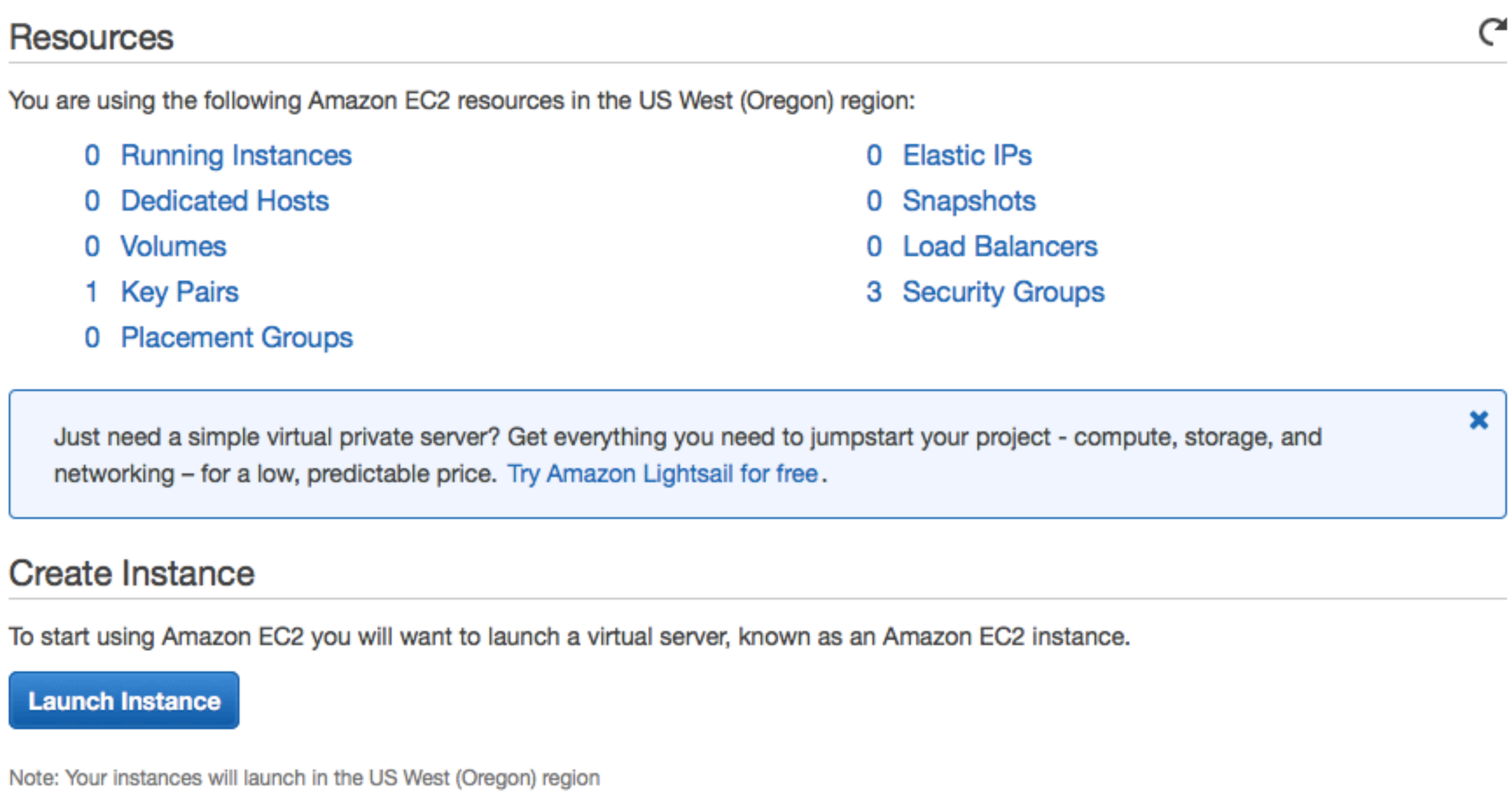

Figure 3.3 Launch Instances (3)

4) The first option "Amazon Linux 64-bit" was selected for this experiment.

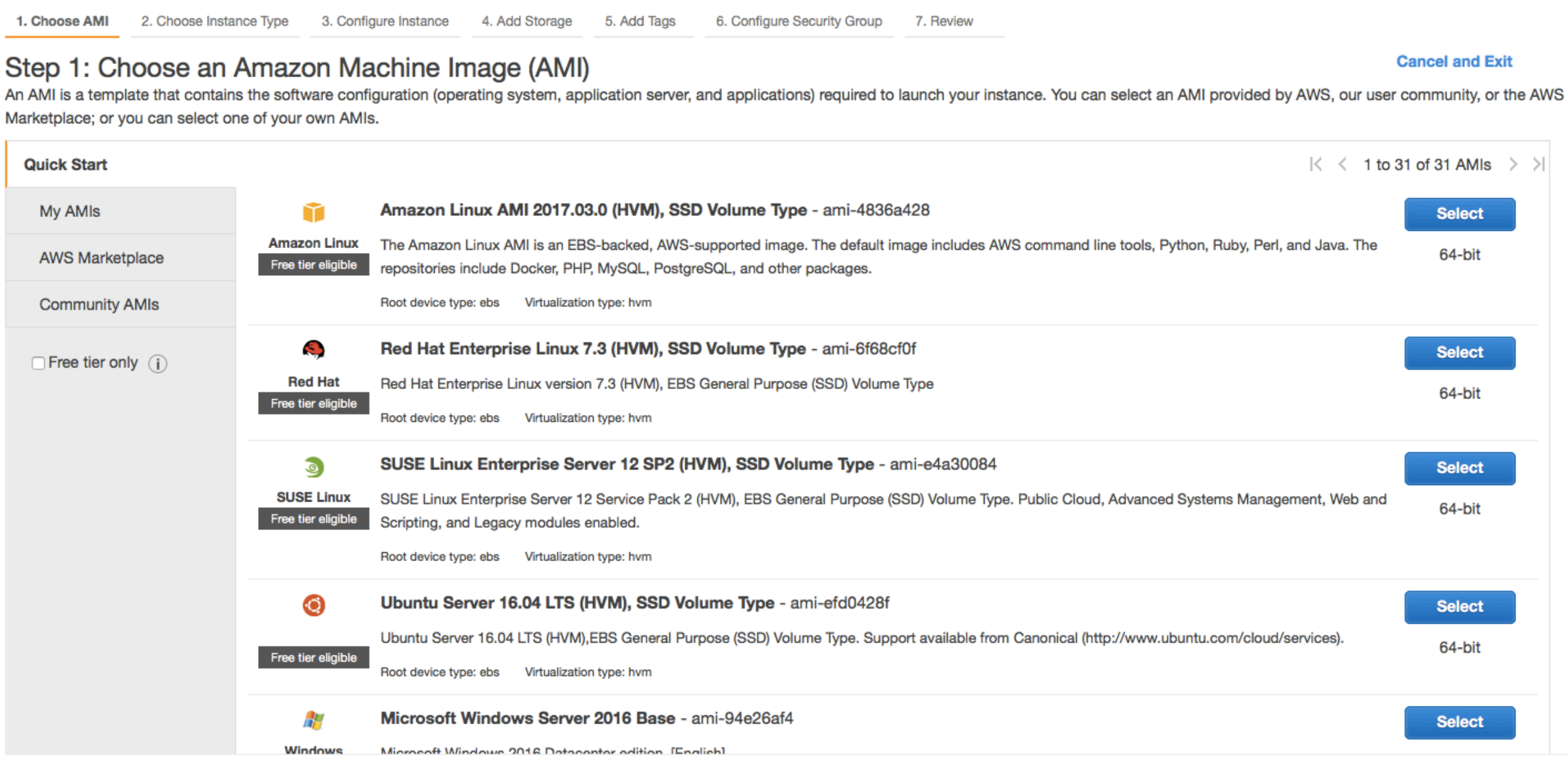

Figure 3.4 Launch Instances (4)

5) Since we have big datasets, so we choose 8 RAM in this case. (all the setup can be done for the free-tier machines as well)

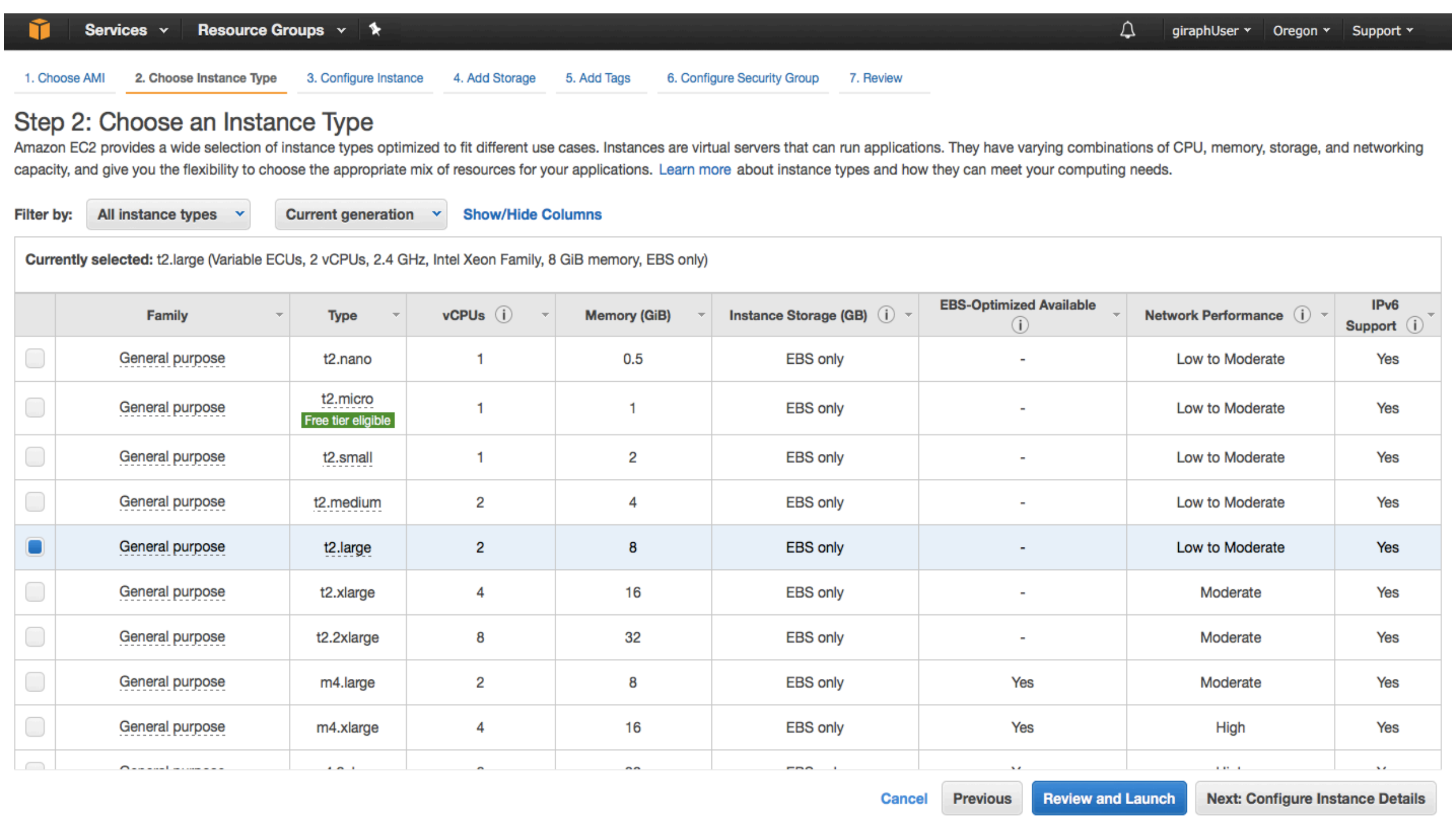

Figure 3.5 Launch Instances (5)

6) Modify the number of instances that we need to launch. Since we need 1 master machine and 20 slaves, we configure 21 instances in total. If the option of specifying the number of instances to launch is not available, one can launch one instance, then press a button for launching more instances like the one just created.

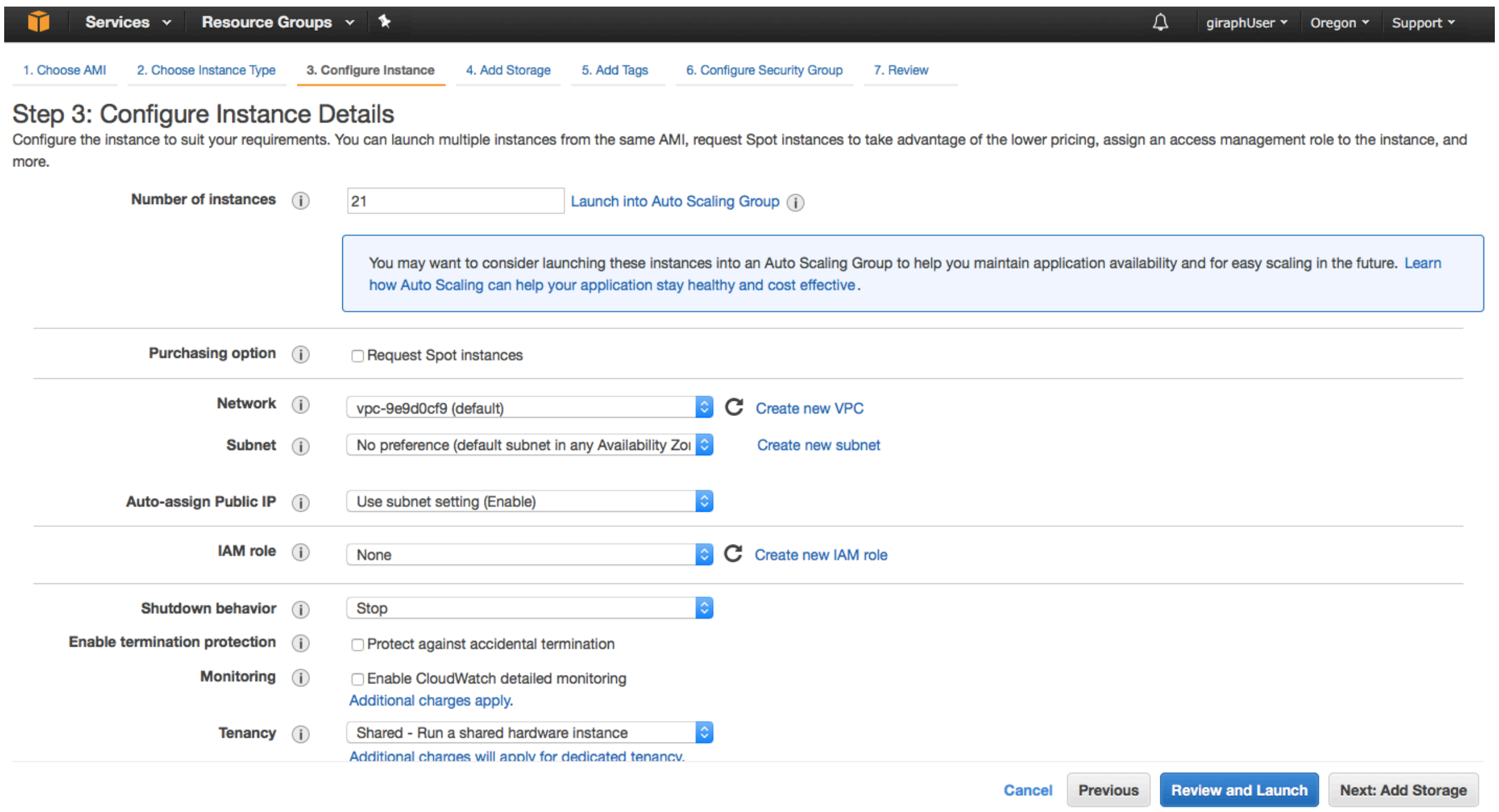

Figure 3.6 Launch Instances (6)

7) Check the detailed information of the instances. If everything is correct, then just click on "Launch" to get started.

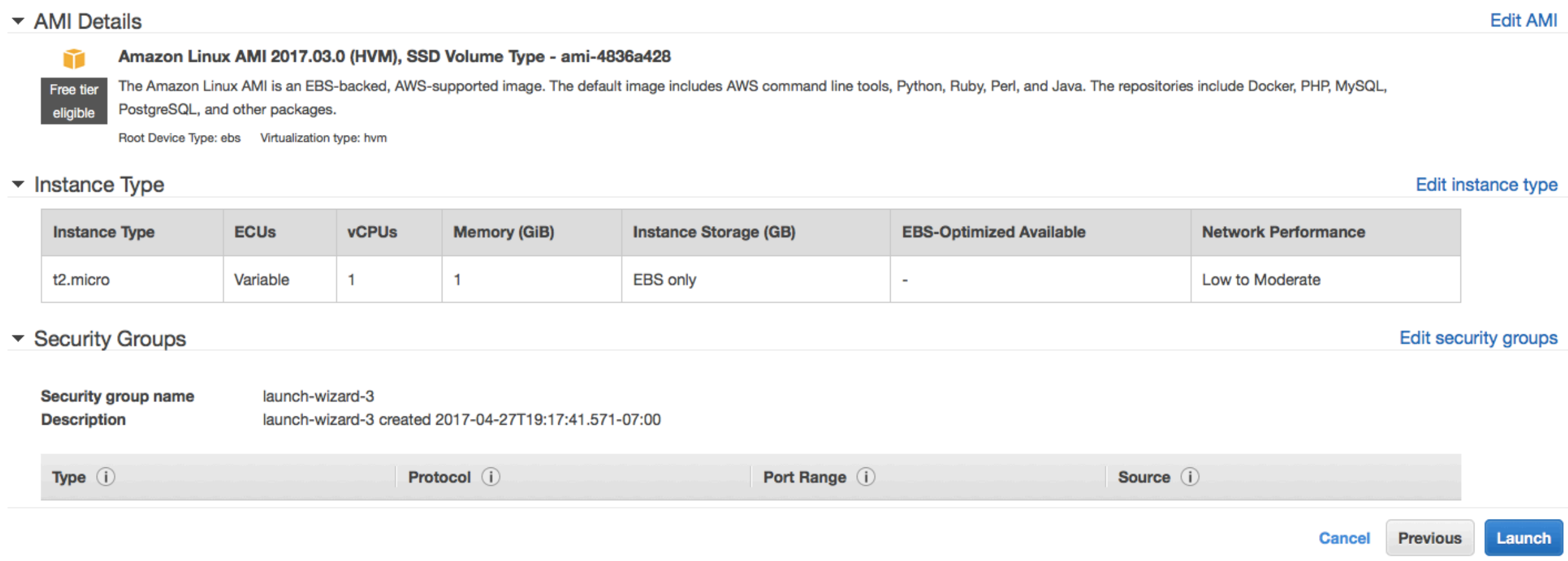

Figure 3.7 Launch Instances (7)

8) We chose "an existing key pair" to launch the instances. If we do not have an existing key pair, then we need to create a new key pair and download the (.pem) file to our local machine.

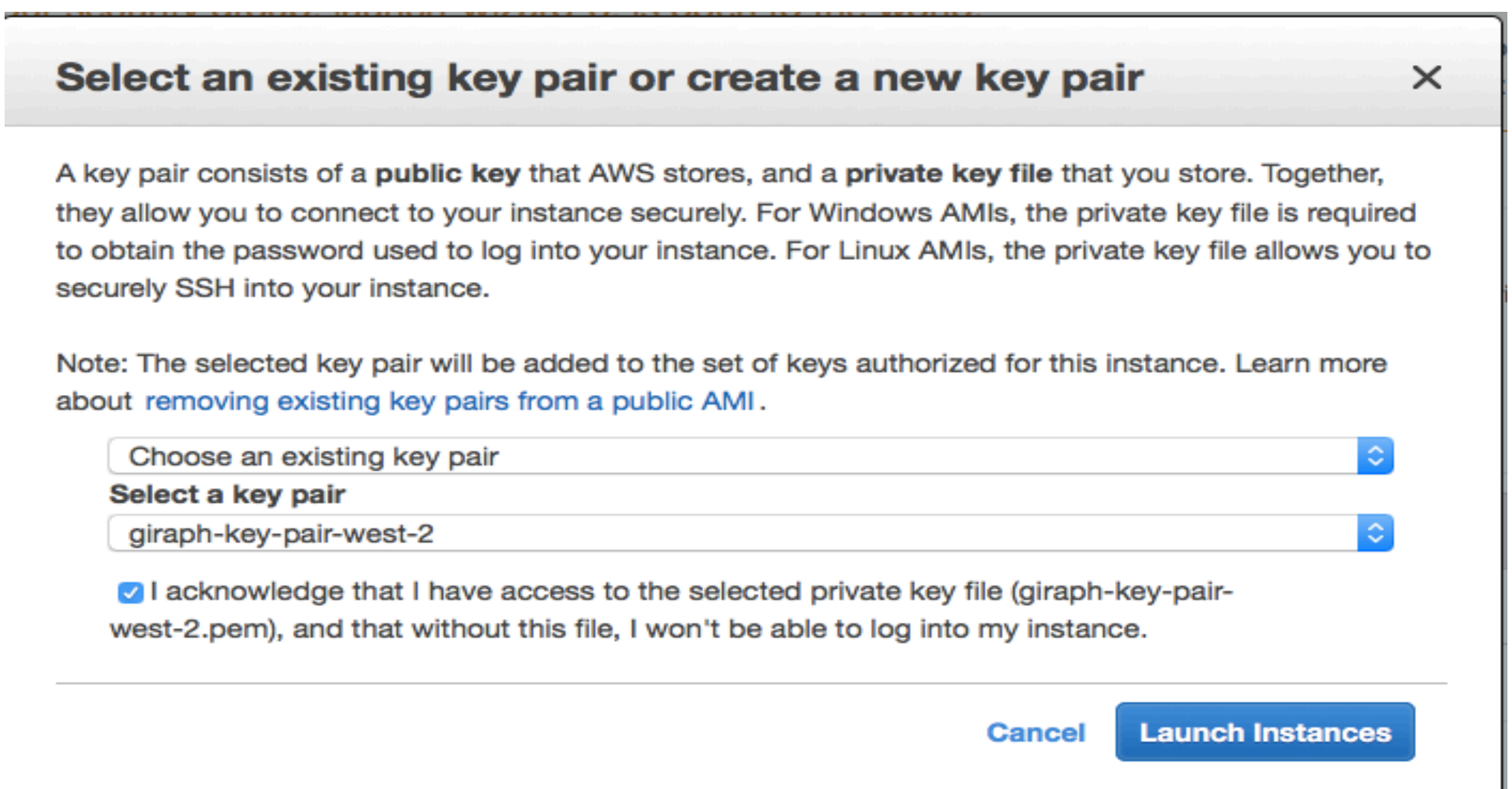

Figure 3.8 Launch Instances (8)

9) Now, we get the confirmation of our instances configurations. Click on "View Instances" to further setup the instances.

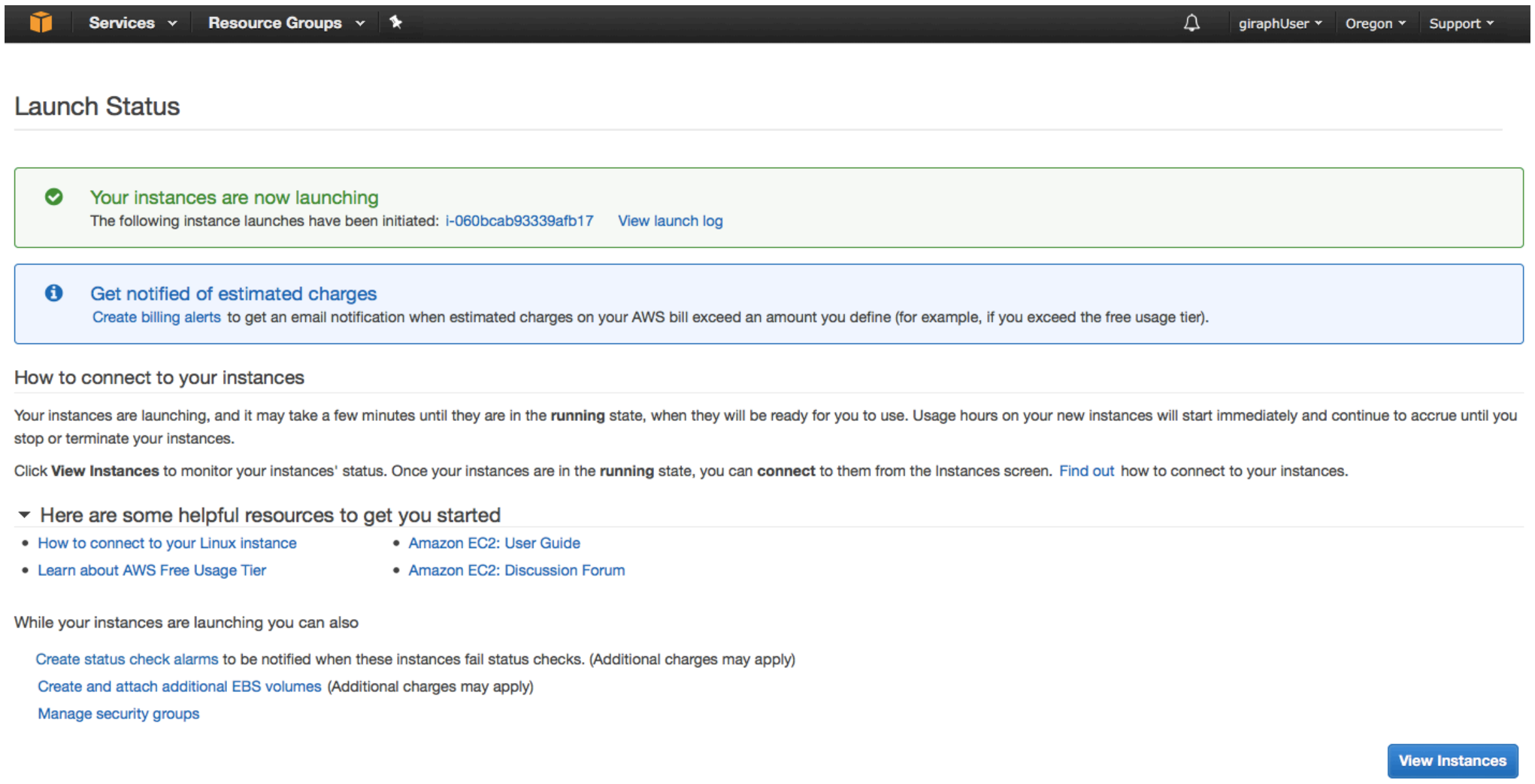

Figure 3.9 Launch Instances (9)

10) Login to each instance, and set up passwordless ssh from the master machine to the slave machines. For details on how to do this please see: https://blog.insightdatascience.com/spinning-up-a-free-hadoop-cluster-step-by-step-c406d56bae42

#### 4.1.2 Environment Setup

1) Update hostname for all instances

$ sudo hostname <Public DNS>

2) Update hostname with its ip address (ifconfig) for all instances

$ sudo vi /etc/hosts

3) Install Java for all instances

$ sudo apt-get update

$ sudo add-apt-repository ppa:webupd8team/java

$sudo apt-get update

$sudo apt-get install oracle-jdk7-installer

4) Download the hadoop package for all instances

$ wget http://apache.mirror.gtcomm.net/hadoop/common/hadoop-1.2.1/hadoop-1.2.1.tar.gz

$ tar -xzvf hadoop-1.2.1.tar.gz

$ mv hadoop-1.2.1 hadoop

5) Add the path shortcut for all instances

$ vi .bashrc

export HADOOP_CONF=/home/ubuntu/hadoop/conf

export HADOOP_PREFIX=/home/ubuntu/hadoop

export JAVA_HOME=/usr/lib/jvm/java-7-oracle

export PATH=$PATH:$HADOOP_PREFIX/bin

6) Add the key pair for master machine only

$ chmod 400 <key pair name>.pem

$ eval 'ssh-agent –s'

$ ssh-add <key pair name>.pem

7) Set JAVA_HOME

```
$ vi $HADOOP_CONF/hadoop-env.sh
```

```
export JAVA_HOME=/usr/lib/jvm/java-oracle
```

8) Edit core-site.xml file for master machine only

```
$ mkdir hdfstmp
```

```
$ vi $HADOOP_CONF/core-site.xml
```

```
<configuration>

<property>
<name>fs.default.name</name>
<value>hdfs://<public-DNS>:8020</value>
</property>

<property>
<name>hadoop.tmp.dir</name>
<value>/home/ubuntu/hdfstmp</value>
</property>

</configuration>
```

9) Edit hdfs-site.xml file for master machine only

```
$ vi $HADOOP_CONF/hdfs-site.xml
```

```
<configuration>

<property>
<name>dfs.replication</name>
<value>3</value>
</property>

<property>
<name>dfs.permissions</name>
<value>false</value>
</property>
```

```
</configuration>
```

10) Edit mapred-site.xml file for the master machine only

```
$ vi $HADOOP_CONF/mapred-site.xml
```

```
<configuration>

<property>
<name>mapred.job.tracker</name>
<value>hdfs://<Public DNS>:8021</value>
</property>

<property>
<name>mapred.child.java.opts</name>
<value>-Xmx4096m</value>
</property>

<property>
<name>mapreduce.job.counters.max</name>
<value>620</value>
</property>

<property>
<name>mapreduce.job.counters.limit</name>
<value>1200</value>
</property>

</configuration>
```

10) Add the Public DNS and SNN for master machine only

```
$ vi $HADOOP_CONF/masters
```

then insert the master node's hostname in that file

11) Add the Public DNS of slaves for master machine only

```
$ vi $HADOOP_CONF/slaves
```

12) Go to $HADOOP_CONF and send with Public DNS

```
$ scp masters slaves Ubuntu@<Public DNS>://home/Ubuntu/hadoop/conf
```

13) Leave the masters file empty for slaves

```
$ vi $HADOOP_CONF/masters
```

14) Add its Public DNS to the slaves file for slaves

```
$ vi $HADOOP_CONF/slaves
```

15) Start hadoop

```
$ hadoop namenode –format

$ cd HADOOP_CONF

$ start-all.sh
```

16) Download Giraph for the master machine only

```
$ sudo apt-get install git

$ sudo apt-get install maven

$ sudo git clone https://github.com/apache/giraph.git

$ sudo chown –R Ubuntu giraph
```

### 4.1.3 Running Giraph on EC2

To execute our algorithm, we execute the following commands:

```
$HADOOP_PREFIX/bin/hadoop dfs -copyFromLocal $HOME/inputFile.txt
/user/ubuntu/input/inputFile.txt

$HADOOP_PREFIX/bin/hadoop
jar $HOME/giraph/giraph-examples/target/giraph-examples-1.3.0-SNAPSHOT-for-hadoop-
1.2.1-jar-with-dependencies.jar
org.apache.giraph.GiraphRunner org.apache.giraph.examples.Kcore
-vif org.apache.giraph.io.formats.JsonLongDoubleFloatDoubleVertexInputFormat
-vip /user/ubuntu/input/input.txt
-vof org.apache.giraph.io.formats.IdWithValueTextOutputFormat
-op /user/ubuntu/output/KcoreOutput -w 1
```

## 5. Algorithm Implementation

The algorithm we consider in this work was initially introduced by Montresor, De Pellegrini and Miorandi in [22] and further engineered by Khaouid, Barsky, Srinivasan, and Thomo in [17]. The algorithm is distributed and follows the "vertex-centric" model of computation. The UDF for Giraph is written in Java. The pseudo code of compute function (Algorithm 1) and computeUpperBound function (Algorithm 2) are respectively shown below.

**Algorithm 1 Compute function running at vertex**

```
function compute(Vertex vertex, Iterable messages)
        if superstep = 0 then
                vertex.value ← vertex.getNumEdges
                sendMessageToAllEdges(vertex)
        else
                localEstimate ← computeUpperBound(vertex, messages)
         if localEstimate < vertex.value then
            vertex.setValue ← localEstimate
            sendMessageToAllEdges(vertex)
         end if
         halt ← true
         for all message in vertex.Messages do
            if vertex.value > massage
               halt ← false
            end if
         end for
         if halt then
            certex.voteToHalt
         end if
```

```
        end if
end function
```

**Algorithm 2 computeUpperBound function for a vertex**

```
function computeUpperBound(Vertex vertex)
    for all i ← 1 to vertex.value do
        c[i] ← 0
    end for
    for all message in vertex.Message do
        j ← min{message, vertex.value}
        c[j] ++
    end for
    cumul ← 0
    for all i ← vertex.value down to 2 do
        cumul ← cumul + c[i]
        if cumul >= i do
            return i
        end if
    end for
end function
```

Algorithm 2 calculates the coreness of each vertex, which can get close to the actual value of the k-core and can effectively speed up the computation of algorithm 1. Algorithm 2 creates an array with size of the vertex degree. For each of the incoming message, algorithm 2 takes the minimum value j between the message value and the current vertex value and then increases the corresponding element by one at the index of j. Adding the elements from the end of the array, once the sum is greater than or equal to its current index i, then i will be finally returned as the coreness of the vertex.

Algorithm 1 initializes the vertex value as the number of its outgoing edges. And the vertex value will be updated to localEstimate if the current value is greater than the coreness of the vertex that computed by algorithm 2. Once the vertex value is the smallest one among its neighbours, the vertex will vote to halt.

To collect average updated times of vertex, the percentage of updated vertices at each superstep, average k-core, and maximum k-core, we used three aggregators to gather our results from each slave at the end of the computation.

## 6. Results and Analysis

Results for the Giraph implementation are shown in Table 4.1. Column "Sent Messages" gives the total numbers of messages that were sent during the whole computation. Column "Update Times" gives the average vertex update times for each dataset. "K-Max" and "K-Ave" are the maximum and average k-core numbers for each dataset. From the table, we can observe that the number of sent messages and vertex update times are not only dependent on the size of the datasets, but also on K-Max and K-Ave. The larger the latter numbers are, the more frequent the message sending and vertex updates will be.

| Dataset Name | \|V\| | \|E\| | Sent Messages | Update Times | K-Max | K-Ave |
|---|---|---|---|---|---|---|
| ca-AstroPh | 18.7 K | 198.1 K | 5,104,983 | 5.414 | 17 | 2.01 |
| p2p-Gnutella31 | 62.6 K | 147.9 K | 322,906 | 0.28 | 50 | 1.143 |
| amazon0601 | 0.4 M | 2.4 M | 12,122,458 | 2.284 | 10 | 2.51 |
| roadNet-CA | 2.0 M | 2.8 M | 11,035,492 | 0.785 | 6 | 1.999 |
| soc-LiveJournal1 | 4.8 M | 43.1 M | 888,141,866 | 3.507 | 434 | 1.689 |

Table 4.1 Giraph computation results

Figure 4.1 shows the number of iterations executed on Giraph and GraphChi. The reason why the iteration numbers for the same dataset are different is that we cannot control the order of running each vertex in distributed cluster-based Giraph. However, because of the selective scheduling feature, the order is fixed when running GraphChi on a single machine. Except for the largest dataset, Giraph needs less iteration than GraphChi with its advantage of running on multiple machines. The percentage of updated nodes over several iterations is shown in Figure 4.2.

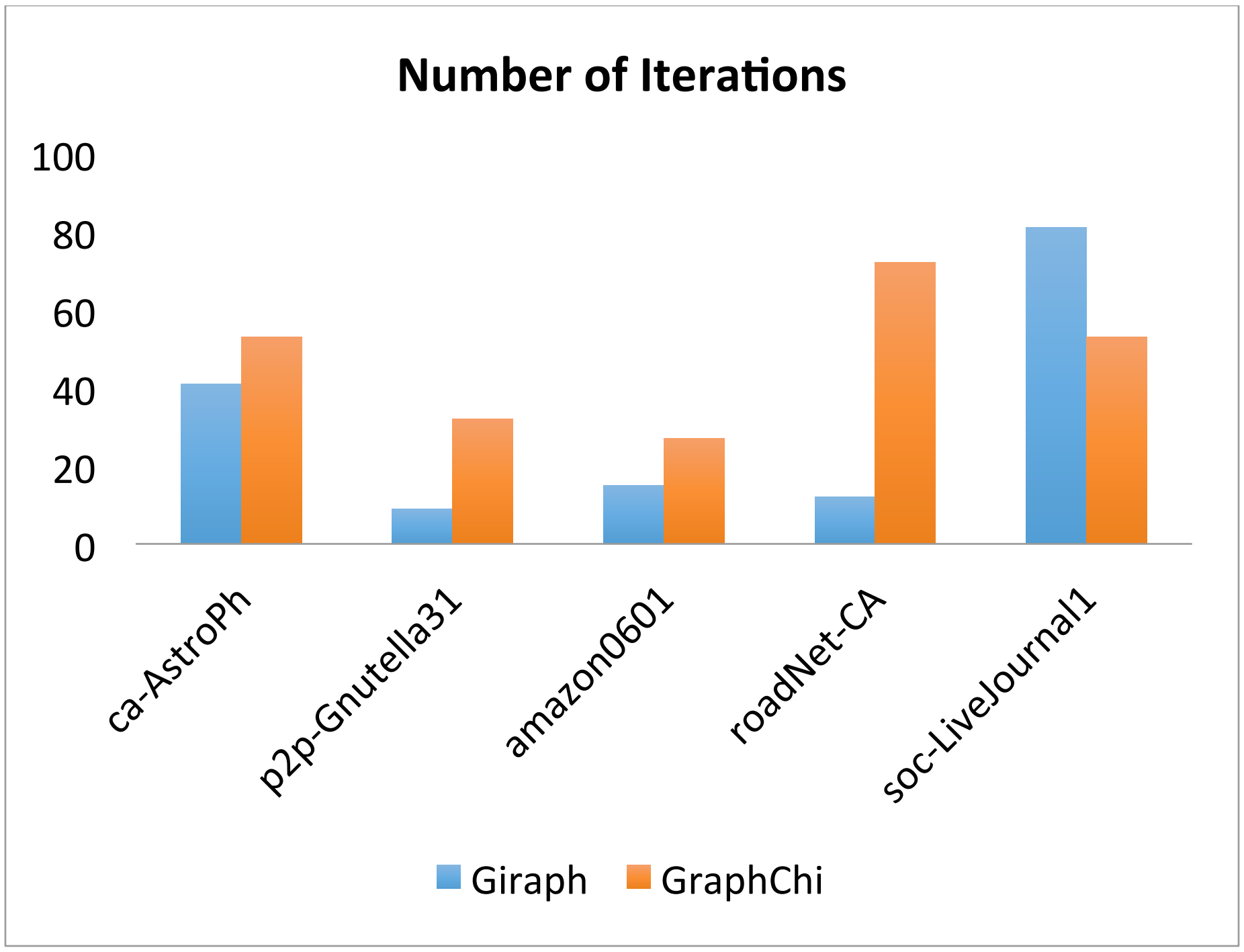

Figure 4.1 Number of iterations

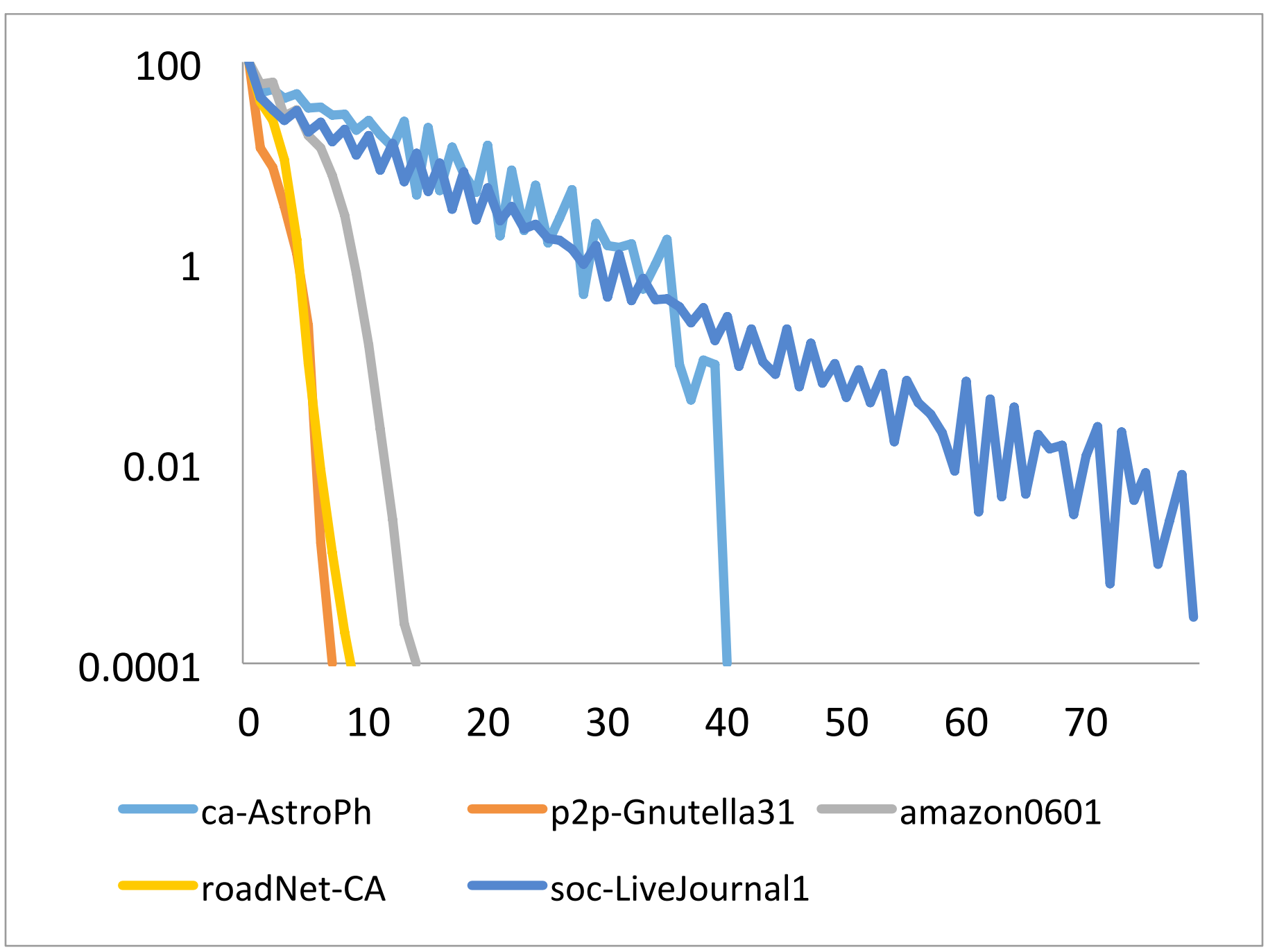

Figure 4.2 Percentage of updated nodes in Giraph vs. number of iterations

Figure 4.4 and 4.5 show the running time (milliseconds) of Giraph versus the number of slave machines used. In Figure 4.4, we see that with the increase in the number of machines, the running time also increased. The more machines we have, the fewer tasks will be assigned to each one. However, the more machines we have, the more time they need to spend on communication. That is why the running time does not decrease when we configure more machines for Giraph. For the largest dataset shown in Figure 4.5, we can notice that Giraph with two slave machines needs the most running time for the computation, which is around 800 seconds. On the contrary, it takes the least running time with ten machines, which can finish the computation within around 700 seconds.

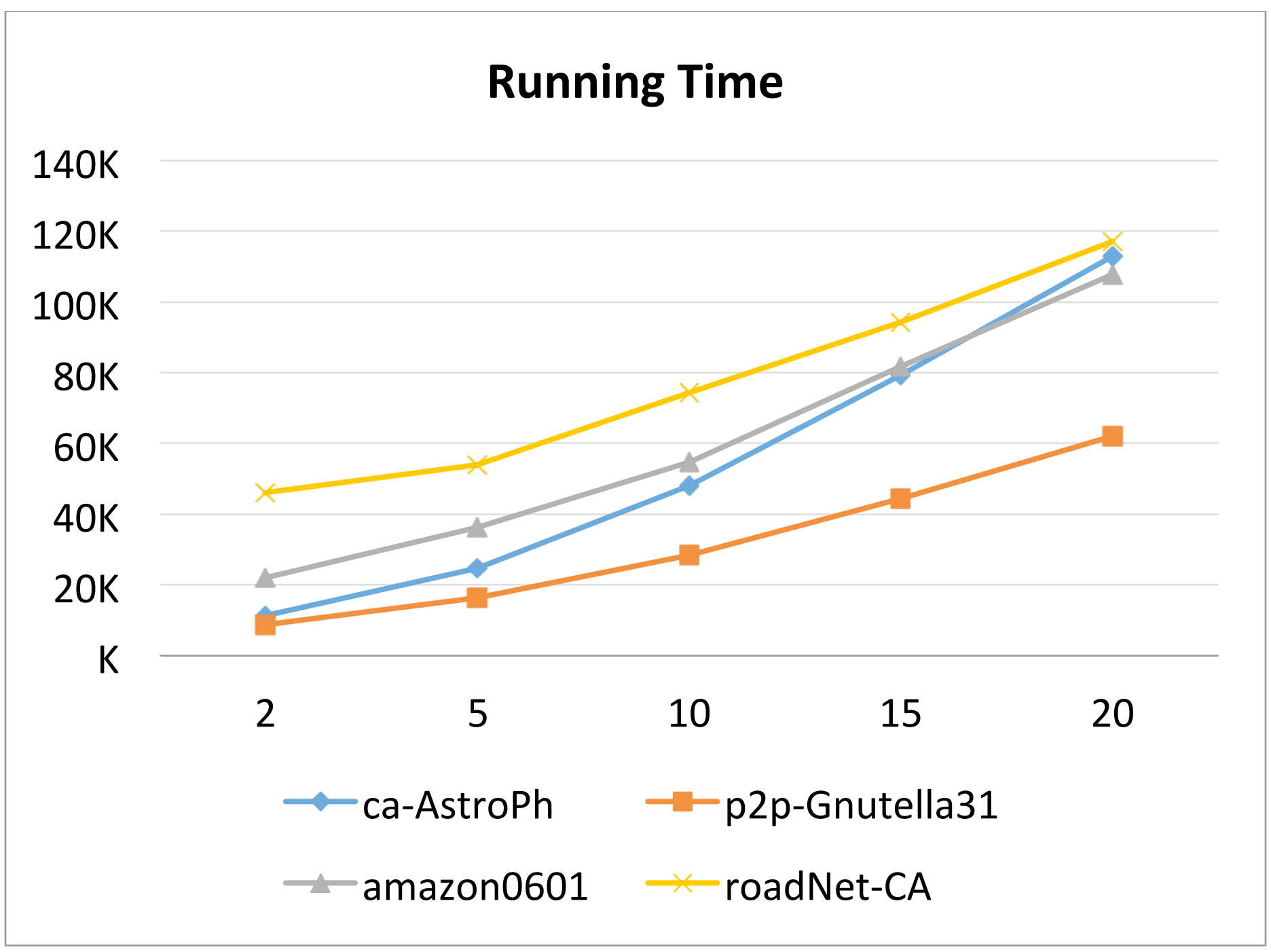

Figure 4.4 Running time (ms) in Giraph vs. number of machines.

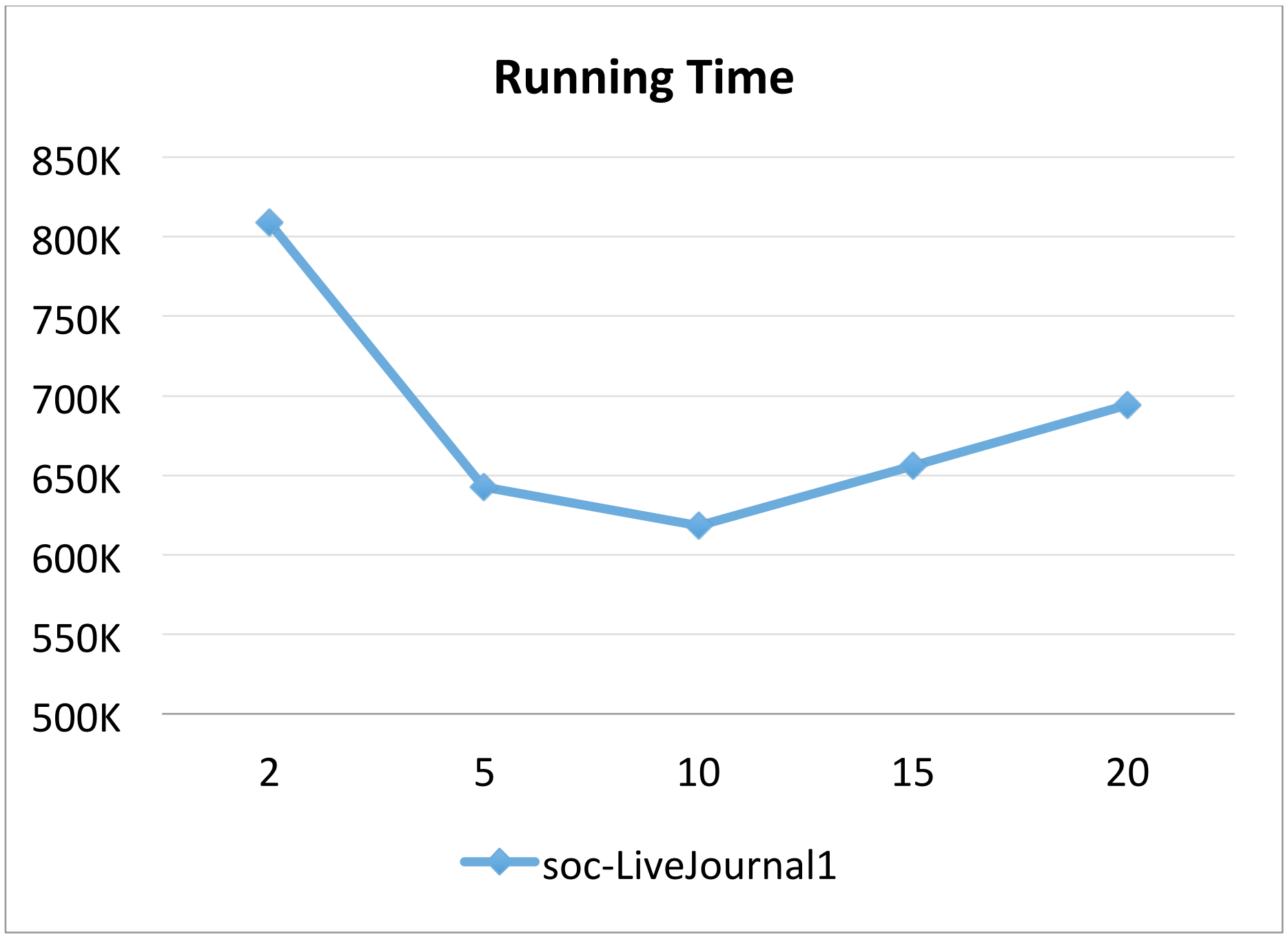

Figure 4.5 Running time (ms) in Giraph vs. number of machines for soc-LiveJournal1

To compare the running time with Giraph and GraphChi, we select the least running time with the proper number of machines for each dataset for Giraph. Figure 4.6 shows the running time comparisons between Giraph and GraphChi. The running time of Giraph and GraphChi are very close. When dealing with the small data ca-AstroPh and p2p-Gnutella31, GraphChi is faster than Giraph. However, Giraph is more efficient to compute k-core for the medium size and large size datasets with the proper number of slaves than GraphChi with a single machine.

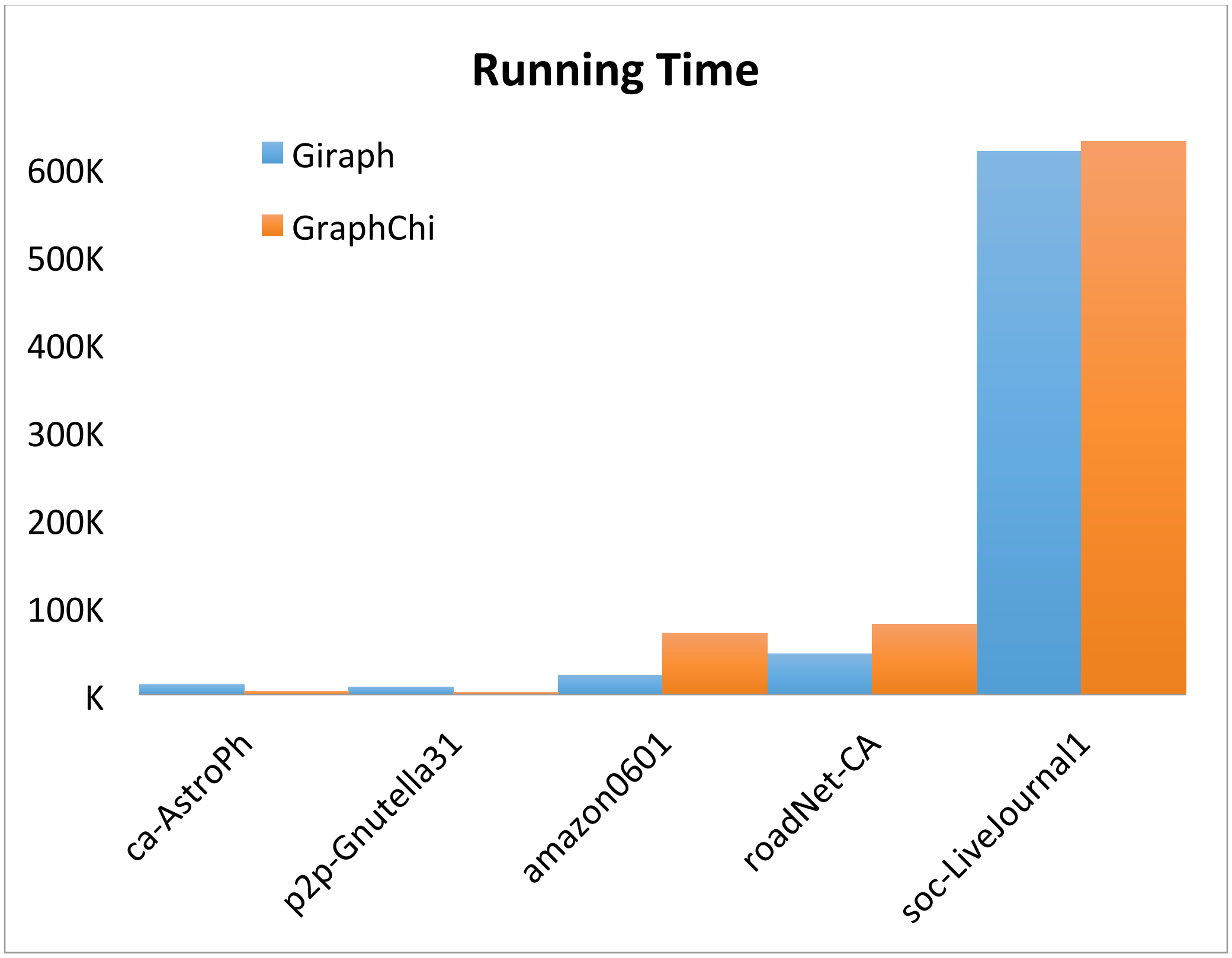

Figure 4.6 CPU running time (ms)

## 7. Related Works

Connections between people or entities are modeled as graphs, where vertices represent the people or entities, and edges represent the connections. Analyzing the graph structure has been shown to be highly beneficial in targeted advertising [33], fraud-detection [23], missing link prediction [21, 18], locating functional modules of interacting proteins [31, 14], identifying new emerging trends in scientific disciplines [5], and so on.

k-core decomposition has many applications. It is extensively used in aiding the visualization of the network structure [3, 25], understanding and interpreting cooperative processes in social networks [12, 11], capturing structural diversity in social contagion [30], analyzing complex networks in terms of node hierarchies, self-similarity, and connectivity [2], describing protein functions based on protein-protein networks [1, 20], exploring collaboration in software teams [32], facilitating network approaches for large text summarization [4], and approximating hard to compute network centrality measures [16]. As future work, we would like to explore the usefulness of k-core decomposition in trust prediction [18], in clearing a contamination from a network [24, 27], in identifying community formation in biological networks [14], and in devising network-based collaborative filtering algorithms [10, 34]. Also, of interest is the extension of the notion of k-core decomposition to probabilistic graphs [7, 15] and to edge-labeled graphs [13, 26].

The algorithm we consider in this work was initially introduced by Montresor, De Pellegrini and Miorandi in [22] and further engineered by Khaouid, Barsky, Srinivasan, and Thomo in [17]. The algorithm is distributed and follows the "vertex-centric" model of computation. It operates on the premise that the input graph is spread across multiple cluster nodes or hosts.

In the case where the large graph resides on a single machine's disk, a single machine framework following the vertex-centric model is GraphChi [19], a modern, general-purpose, graph engine which employs a novel technique for processing large data from disk. Experiments for k-core decomposition using GraphChi are presented in [17]. When the graph does not fit in main memory (but fits in disk), a well-known algorithm for k-core decomposition of massive networks is EMcore proposed by Cheng, Ke, Chu, and Ozsu in [9].

As future work, we would like to compare the Giraph distributed framework to the distributed setting of algorithms in [26, 29]. Also of interest is to use graph compression frameworks, such as Webgraph [6], to be able to handle much larger graphs, e.g. [8, 27, 28].

## 8. Conclusions

With the experiments of k-core computation, we observe that Giraph is suitable for analyzing large data since it can synchronously implement the computation by assigning the tasks to each slave. However, it is not very fast to compute on small data, because slaves need time to communicate with each other. Conversely, GraphChi implemented on a single machine can avoid the communication overheads. It is more efficient than Giraph for computing on small data.